Data base

# VenusMutHub: A systematic evaluation of protein mutation effect predictors on small-scale experimental data

**Liang Zhang[a,d†], Hua Pang[b,c†], Chenghao Zhang[e†], Song Li[a], Yang Tan[a,c,d], Fan Jiang[a,d], Mingchen Li[a,c,d], Yuanxi Yu[a,d], Ziyi Zhou[a,d], Banghao Wu[a,e], Bingxin Zhou[a], Hao Liu[f], Pan Tan[a,c,d*], Liang Hong[a,d,e*]**

[a]*School of Physics and Astronomy & Institute of Natural Sciences, Shanghai Jiao Tong University, Shanghai National Centre for Applied Mathematics (SJTU Center), MOE-LSC, Shanghai 200240, China*

[b]*iHuman Institute and School of Life Sciences and Technology, Shanghaitech University, Shanghai 201210, China*

[c]*Shanghai Artificial Intelligence Laboratory, Shanghai 200232, China*

[d]*Zhangjiang Institute for Advanced Study, Shanghai Jiao Tong University, Shanghai 201203, China*

[e]*School of Life Sciences and Biotechnology, Shanghai Jiao Tong University, Shanghai 200240, China*

[f]*Shanghai Matwings Technology Co., Ltd., 200240 Shanghai, China*

*Corresponding authors.

E-mail addresses: tpan1039@gmail.com (Pan Tan), hongl3liang@sjtu.edu.cn (Liang Hong).

[†]These authors made equal contributions to this work.

Running title: Small-scale mutation benchmark

**Abstract** In protein engineering, while computational models are increasingly used to predict mutation effects, their evaluations primarily rely on high-throughput deep mutational scanning (DMS) experiments that use surrogate readouts, which may not adequately capture the complex biochemical properties of interest. Many proteins and their functions cannot be assessed through high-throughput methods due to technical limitations or the nature of the desired properties, and this is particularly true for the real industrial application scenario. Therefore, the desired testing datasets, will be small-size (~10-100) experimental data for each protein, and involve as many proteins as possible and as many properties as possible, which is, however, lacking. Here, we present VenusMutHub, a comprehensive benchmark study using 905 small-scale experimental datasets curated from published literature and public databases, spanning 527 proteins across diverse functional properties including stability, activity, binding affinity, and selectivity. These datasets feature direct biochemical measurements rather than surrogate readouts, providing a more rigorous assessment of model performance in predicting mutations that affect specific molecular functions. We evaluate 23 computational models across various methodological paradigms, such as sequence-based, structure-informed and evolutionary approaches. This benchmark provides practical guidance for selecting appropriate prediction methods in protein engineering applications where accurate prediction of specific functional properties is crucial.

1.Introduction

Protein engineering has emerged as a powerful approach for developing proteins with enhanced or novel functions, playing crucial roles in various applications ranging from industrial biocatalysis to therapeutic protein development[1–3]. The significance of mutation effect prediction is particularly profound in pharmaceutical science, where understanding protein mutations is essential for drug development and precision medicine[4]. In biopharmaceutical development, accurate prediction of mutation effects can guide the optimization of protein therapeutics for improved stability, manufacturability, and clinical efficacy[5,6]. Moreover, in precision medicine, understanding the impact of missense mutations in drug targets is crucial for predicting patient-specific drug responses and developing targeted therapeutic strategies[7]. However, traditional wet-lab experimental methods are often limited by their throughput and resource requirements when screening the vast mutational sequence space, making it impractical to experimentally test all possible mutations. This limitation has highlighted the critical role of computational approaches in protein engineering. A central challenge in this field is the accurate prediction of mutation effects, which holds the potential to significantly accelerate protein design workflows by prioritizing promising candidates for experimental validation[8, 9].

Recent years have witnessed remarkable progress in zero-shot computational methods for predicting protein mutation effects. These approaches span from physics-based methods[10] to modern machine learning models[11], offering diverse strategies for mutation effect prediction. Deep learning models have shown significant advancements in predicting the effects of protein mutations, leveraging their capabilities in understanding complex sequence-function relationships[12]. However, most of these models have been primarily evaluated on large-scale mutation datasets, which derived from deep mutational scanning experiments[13]. While such evaluation provides valuable insights into overall model performance, they have a fundamental limitation: to achieve high-throughput screening, DMS experiments must rely on approximate assays and surrogate readouts (*e.g.*, fluorescence-based assays or growth-based selections) that may not adequately capture or reflect the actual biochemical properties of interest[14, 15]. This limitation becomes particularly critical when the target properties involve complex molecular functions or require specific physiological contexts that can hardly be approximated by high-throughput screening methods.

In real-world protein engineering scenarios, particularly in directed evolution applications, researchers often employ an iterative approach where initial experimental data from a small number of carefully selected mutations guide subsequent rounds of engineering. These mutations are typically chosen from functionally important regions based on structural knowledge or mechanistic understanding. Many proteins of interest cannot be subjected to high-throughput screening methods due to technical limitations, cost constraints, or the nature of the desired function[16]. Consequently, available experimental data often consists of small-scale datasets, typically ranging from dozens to several hundred mutations. The first-round experimental data not only validates initial predictions but also provides crucial information for selecting the most suitable prediction model for the next round of engineering. This creates a significant gap between the evaluation conditions of prediction models and their practical application contexts, where accurate prediction of mutations in key regions is essential for efficient iterative optimization.

Furthermore, protein engineers need to predict various functional properties including stability, catalytic activity, binding affinity and selectivity. Different prediction models may exhibit varying performance across these properties, and their effectiveness might depend on the size and quality of available training data. Given these challenges, the evaluation of models on small-scale datasets is particularly crucial because: (1) it reflects the real constraints in industrial and academic protein engineering projects[16], (2) it enables assessment using direct biochemical measurements of the target properties rather than surrogate readouts[14], and (3) it provides insights into model selection for the next round of engineering based on initial experimental data. Despite the abundance of prediction methods, there lacks a systematic evaluation of their performance on small-scale datasets that better represent real-world protein engineering scenarios through rigorous experimental characterization.

Motivated by these practical challenges, we present VenusMutHub, a comprehensive benchmark study evaluating diverse mutation effect prediction models. We systematically collected and curated 905 small-scale experimental mutation datasets from published literature and public databases, spanning 527 proteins across various functional properties, including stability, activity, binding affinity, and selectivity. Our evaluation encompasses diverse prediction approaches including protein language models, structure-aware models, and alignment-based methods, assessing their performance across different dataset sizes and functional properties. This benchmark aims to provide practical insights for protein engineers by objectively assessing model performance under realistic economic and technical constraints.

## 2. Materials and methods

### 2.1. Dataset construction

**Data collection** We compiled our benchmark datasets from published literature and public databases, with a focus on protein engineering applications. For stability measurements ($\Delta\Delta G$, $\Delta T_m$), we integrated several established datasets including FireProtDB and ThermoMutDB[17, 18], along with other curated stability datasets (S571, S4346, S8754, M1261[19], S669[20], S2648[21]). For binding affinity data, we curated two types of datasets: protein–protein interaction (PPI) data from PPB-Affinity[22], focusing on single-point mutations occurring on a single chain, and drug-target interaction (DTI) data from BindingDB[23], including $K_d$, $K_i$, and $IC_{50}$ measurements for mutations of the same wild-type protein. We also curated catalytic activity ($k_{cat}$, $k_{cat}/k_m$, $k_m$, specific activity and $v_{max}$) and selectivity (e.e., d.e.) data from published protein engineering studies. To ensure data quality, we only included datasets with clear documentation of experimental methods and quantitative measurements.

**Dataset characteristics** As shown in Fig. 1A, our benchmark encompasses 905 distinct datasets from 527 unique proteins, with stability measurements constituting the majority (59.7%, 540 datasets), followed by activity (19.3%, 175 datasets), binding (15.8%, 143 datasets), and selectivity (5.2%, 47 datasets). The protein sequence length distribution (Fig. 1B) spans from small peptides to large proteins (0–1400 residues). The distribution of dataset sizes (Fig. 1C) reflects typical protein engineering scenarios, where most experiments involve a limited number of mutations: 578 datasets are very small (5–20 mutations), 197 are small (21–50 mutations), 77 are medium-sized (51–100 mutations), and 53 datasets contain more than 100 mutations. This size distribution aligns well with real-world protein engineering practices, where comprehensive deep mutation scanning is often impractical.

**Data processing** To ensure data quality and consistency, we implemented a comprehensive preprocessing pipeline. We applied strict quality control criteria by removing mutations where the annotated wild-type residue did not match the corresponding position in the reference sequence (*e.g.*, a mutation annotated as M1H was removed if the first residue in the wild-type sequence was not methionine). Additionally, we required each dataset to contain at least 5 mutations with corresponding experimental measurements to ensure sufficient data points for analysis. To address potential redundancy, we developed a deduplication process using MD5 hash values based on mutation and score information. The processed datasets were standardized to a consistent format containing mutant identifiers, mutation sequences, and experimental measurements, facilitating uniform model evaluation across all datasets.

For structural information required by various prediction models, we utilized ColabFold[24] to generate both multiple sequence alignments (MSAs) and protein structure predictions. The MSAs generated by ColabFold in a3m format were converted to a2m format using the reformat.pl script from HH-suite3[25] to accommodate the input requirements of various prediction models. This standardized pipeline ensured consistent structural and evolutionary information across all proteins in our benchmark.

### 2.2. Evaluated models

We evaluated a comprehensive set of 23 protein mutation effect prediction models, which can be categorized based on their input features:

### 2.2.1. Sequence-only models

These models only require the target protein sequence as input:

**ESM** ESM-1b[11], ESM-1v[26] and ESM-2[27] are protein language models with transformer encoder architecture similar to BERT, trained with masked language modeling objectives on UniRef50 or UniRef90. They predict fitness using the masked-marginal approach, which provides optimal performance on substitutions.

**VESPA** VESPA[28] is a Single Amino acid Variant (SAV) effect predictor that combines embeddings from the protein language model ProtT5 with per-residue conservation predictions.

**CARP** CARP[29] is a protein language model trained with MLM objective on UniRef50. The architecture leverages convolutions instead of self-attention, leading to computational speedups while maintaining high downstream task performance.

**RITA** RITA[30] is an autoregressive language model akin to GPT2[31], trained on UniRef100 with model sizes ranging from 85M to 1.2B parameters.

**ProGen2** ProGen2[32] is an autoregressive protein language model trained on a mixture of UniRef90 and BFD30[33], with models ranging from 151M to 6.4B parameters. It employs a standard transformer decoder architecture.

**ProtGPT2** ProtGPT2[34] is a 738M-parameter autoregressive transformer trained on Uniref50, capable of generating de novo protein sequences that explore novel regions of the protein space while maintaining natural protein-like properties.

**UniRep** UniRep[35] trains a Long Short-Term Memory (LSTM) model on UniRef50 sequences.

### 2.2.2. Evolution-informed models

These models leverage evolutionary information through different approaches:

**GEMME** GEMME[36] is a Global Epistatic Model that predicts mutational effects by analyzing multiple sequence alignments. It combines evolutionary tree-based conservation analysis with site-wise frequencies to calculate epistatic effects across the entire sequence, providing a statistical approach to mutation effect prediction.

**MSA transformer** MSA Transformer[37] directly processes multiple sequence alignments using a specialized transformer architecture designed to capture dependencies both within and across sequences in the alignment.

**Tranception** Tranception[38] offers flexibility in utilizing MSA information through its autoregressive architecture, with variants available for both MSA-based and sequence-only prediction.

**PoET** PoET[39] takes an innovative approach by modeling protein families using unaligned homologous sequences with an autoregressive architecture.

**VenusREM** VenusREM[40] integrates sequence, structure, and evolutionary information through a unified transformer architecture. It processes protein structures via a structure tokenization module, while incorporating evolutionary information through an Alignment Tokenization Module, and combines these features using disentangled multi-head attention for mutation effect prediction.

### 2.2.3. Structure-aware models

These models incorporate structural information in different ways:

**MULAN** MULAN[41] builds upon ESM-2 architecture by adding a Structure Adapter module to incorporate backbone angle information, enabling effective integration of structural features while maintaining the power of pretrained language models.

**ProSST** ProSST[12] is a structure-aware protein language model that consists of two key components: a GVP[42]-based structure quantization module that encodes local residue environments into discrete tokens, and a transformer with sequence-structure disentangled attention that explicitly models the relationship between sequence and structural features. The model is pre-trained on 18.8 million protein structures using masked language modeling.

**ProtSSN** ProtSSN[43] combines semantic and geometric encoding to capture both sequence and structural information. It uses ESM-2 for semantic encoding of protein sequences and Equivariant Graph Neural Networks (EGNN)[44] for geometric encoding of protein structures. The model is trained through self-supervised learning by recovering protein information from noisy sequences and structures.

**SaProt** SaProt[45] introduces a structure-aware vocabulary that combines residue and 3D geometric information through Foldseek[46]-based structure tokenization. This 650M-parameter model converts protein sequences into structure-aware tokens that can be processed by standard language model architectures.

**Inverse folding models** MIFST[47], MIF[47], ProteinMPNN[48] and ESM-IF1[49] are inverse folding models specifically designed to model the relationship between protein sequence and structure, generating sequences compatible with given backbone structures.

All models were evaluated using their publicly available implementations and pretrained weights. For models requiring structural information in single-chain prediction tasks, we utilized protein structures generated by ColabFold[50]. For protein–protein interaction (PPI) predictions, ESM-IF1 and ProteinMPNN were evaluated using experimental complex structures from the PDB database, as these models can process multi-chain inputs. Multiple sequence alignments, when required, were also generated using the ColabFold pipeline to ensure consistency across evaluations.

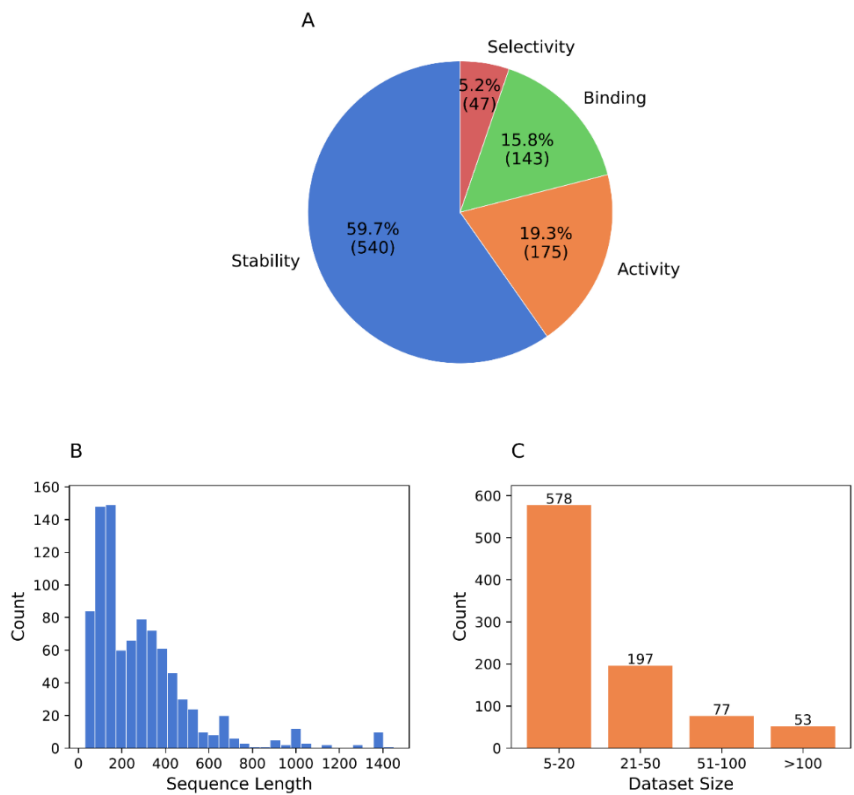

**Figure 1** Overview of the dataset distribution. (A) Distribution of protein property categories in the dataset, showing the percentage and count for each category. Stability includes measurements like $\Delta\Delta G$, which represents the change in free energy difference between a protein's folded and unfolded states, and $\Delta T_m$, which indicates the change in melting temperature, showing how resistant a protein is to unfolding when heated. Activity encompasses catalytic measurements including $k_{cat}$, $k_{cat}/k_m$, $k_m$, specific activity and $v_{max}$; Binding refers to protein–protein interaction (PPI) affinities and drug-target interaction (DTI) measurements ($K_d$, $K_i$, $IC_{50}$); Selectivity represents stereoselectivity measurements (e.e., d.e.). (B) Distribution of protein sequence lengths across the dataset, binned into 30 equal intervals for visualization. (C) Distribution of dataset sizes based on the number of mutations per protein. The datasets are grouped into four categories (5–20, 21–50, 51–100, and >100 mutations) to reflect the scale of mutation studies, ranging from mutations of specific functional sites to comprehensive exploration of protein functional domains.

## 2.3. Evaluation metrics

To comprehensively assess model performance, we employed four key evaluation metrics that capture different aspects of prediction accuracy. The primary metric we used is the

Spearman correlation coefficient ($\rho$) between predicted and experimental values as shown in Eq. (1):

$$\rho = 1 - \frac{6\sum_{i=1}^{n} d_i^2}{n(n^2 - 1)} \quad (1)$$

where $d_i$ is the difference between the ranks of corresponding predicted and experimental values, and $n$ is the number of mutations. Spearman correlation evaluates models' ability to correctly rank mutations regardless of the absolute scale, which is crucial for prioritizing mutations in protein engineering applications. To assess models' ability to identify the most promising mutations, we calculated the Normalized Discounted Cumulative Gain (NDCG) with the top 20% of mutations as the relevant set as shown in Eqs. (2) and (3):

$$\text{NDCG} = \frac{\text{DCG}}{\text{IDCG}} \quad (2)$$

$$\text{DCG} = \sum_{i=1}^{p} \frac{2^{\text{rel}_i} - 1}{\log_2(i + 1)}, IDCG = \sum_{i=1}^{p} \frac{2^{\text{rel}_i^*} - 1}{\log_2(i + 1)} \quad (3)$$

where $\text{rel}_i$ represents the relevance score calculated as follows: first, mutations are ranked based on model predictions; for mutations ranked in the top 20% by model predictions, their relevance scores are calculated as their experimental values normalized to [0,1] using min-max normalization as shown in Eq. (4):

$$\text{rel}_i = \frac{v_i - v_{\min}}{v_{\max} - v_{\min}} \quad (4)$$

where $v_i$ is the experimental value of mutation $i$, and $v_{\min}$ and $v_{\max}$ are the minimum and maximum experimental values across all mutations. For mutations not in the top 20%, their relevance scores are set to 0. $\text{rel}_i^*$ represents these same relevance scores sorted in descending order (the ideal ranking), and $p$ is the total number of mutations. NDCG is particularly suitable for evaluating mutation prioritization as it measures how well a model's top predictions align with their actual experimental improvements.

For binary classification metrics, we established the classification threshold based on available data characteristics. For datasets containing wild-type fitness measurements, mutations with fitness scores higher than the wild-type were classified as positive samples, while those lower were classified as negative samples. For datasets without wild-type measurements, we used the median fitness score as the threshold. Similarly, predictions were classified as positive if their predicted scores exceeded the corresponding threshold (wild-type prediction or median predicted score). Based on these criteria, we calculated accuracy as shown in Eq. (5):

$$\text{Accurarcy} = \frac{\text{TP} + \text{TN}}{\text{TP} + \text{TN} + \text{FP} + \text{FN}} \quad (5)$$

The F1 score provides a balanced measure between precision and recall, which is particularly important when beneficial mutations are rare as shown in Eqs. (6) and (7):

$$\text{Precision} = \frac{\text{TP}}{\text{TP} + \text{FP}}, \text{Recall} = \frac{\text{TP}}{\text{TP} + \text{FN}} \qquad (6)$$

$$\text{F1} = 2 \cdot \frac{\text{Precision} \cdot \text{Recall}}{\text{Precision} + \text{Recall}} \qquad (7)$$

These complementary metrics provide a comprehensive evaluation: Spearman correlation captures the overall ranking ability, NDCG focuses on identifying top mutations, while accuracy and F1 score assess the models' capability in distinguishing beneficial mutations from deleterious ones.

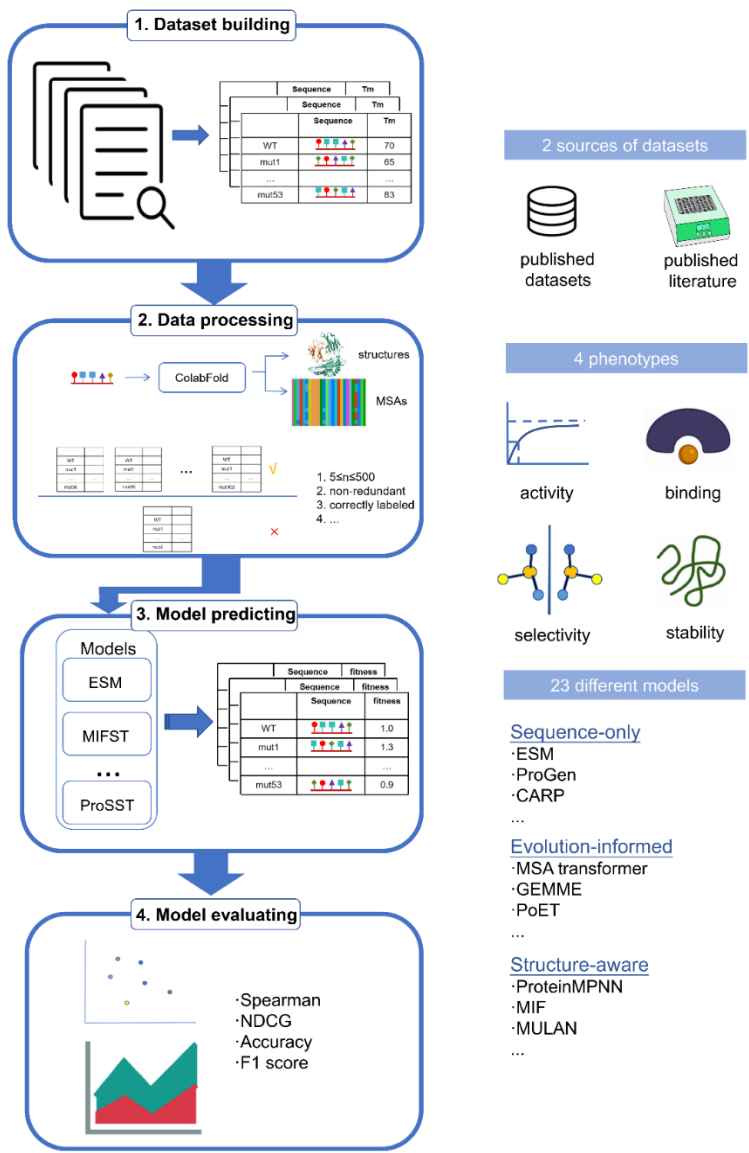

**Figure 2** Overview of the benchmark workflow. Our pipeline consists of four main steps: (1) Dataset building from published datasets and literature; (2) Data processing with ColabFold for structure and MSA generation; (3) Model evaluation across 23 different models (Sequence-only, evolution-informed, and structure-aware); (4) Performance analysis using correlation metrics (Spearman, NDCG) and classification metrics (Accuracy, F1 score). The right panel illustrates the key components of our benchmark: data sources, phenotypes and model categories.

## 3. Results and discussion

### 3.1. Overall performance analysis

We constructed a comprehensive mutation benchmark dataset and systematically evaluated the performance of 23 zero-shot protein mutation fitness prediction models, revealing several key insights about their predictive capabilities (Table 1). VenusREM, which integrates sequence, structure, and MSA information, achieves the highest Spearman correlation and NDCG on both datasets (0.230 and 0.863 for all mutations; 0.285 and 0.878 for single-point mutations). Meanwhile, MIF emerges as another top performer, achieving the highest scores in accuracy and F1 metrics (0.612 and 0.547 for all mutations; 0.627 and 0.562 for single-point mutations), while maintaining competitive performance in ranking metrics (second highest Spearman correlation of 0.229 for all mutations and 0.261 for single-point mutations).

Structure-aware models, particularly inverse folding models (MIF, MIFST, ESM-IF1) and other structure-based approaches (VenusREM, ProSST), consistently rank among top performers in our evaluation. This apparent advantage of structure-aware models, however, should be interpreted with caution as it may be partially attributed to the composition of our dataset, where stability-related mutations constitute a significant portion. We will examine this potential bias in subsequent sections through detailed functional analysis. The strong performance of VenusREM further suggests that the integration of multiple information sources (sequence, structure, and evolutionary information) can provide complementary signals for mutation effect prediction. In particular, we observe distinct performance patterns between the Spearman correlation and NDCG metrics. While VenusREM achieves the highest Spearman correlation (0.285) and NDCG (0.878) for single-point mutations, likely benefiting from its comprehensive information integration, MIF shows exceptional performance in accuracy (0.627) and F1 score (0.562). These variations suggest that different models may excel at capturing different aspects of the mutation landscape.

To ensure the reliability of our analysis, we constructed two datasets: a complete dataset containing both single and multiple mutations (comprising 905 individual datasets), and a refined single-point mutation dataset (comprising 796 individual datasets) by removing all multiple-mutation entries and retaining only those assays with at least 5 single-point mutations. Notably, all models show substantially better performance on the single-point mutation dataset compared to the complete dataset, revealing a universal challenge in capturing epistatic effects. This challenge stems from several fundamental limitations in current modeling approaches. First, the current transformer-based architectures typically predict multiple mutation effects through simple combination of individual mutation impacts, which does not account for non-additive epistatic interactions where one mutation can dramatically alter the structural and functional context for subsequent mutations. Second, the vast combinatorial space of multiple mutations means that most possible combinations are not represented in training data, making it difficult for models to learn generalizable patterns of mutation interactions. These limitations pose significant challenges for protein engineering applications, as multiple simultaneous mutations are often required for achieving desired functional improvements. Therefore, to ensure reliable model assessment, our subsequent analyses focus exclusively on single-point mutations, while acknowledging that improving multiple-mutation predictions remains a critical area for future development. Future advances may come from integrating molecular dynamics simulations to capture long-range epistatic interactions mediated through networks of correlated protein dynamics, or from developing specialized architectures that can explicitly model the cross-correlations between mutation sites Table 1: Overall Model Performance (905 datasets for all mutations and 796 datasets for single-point mutations. The metrics for each dataset are calculated and the values represent the average across all datasets).

Table 1: Overall Model Performance (905 datasets for all mutations and 796 datasets for single-point mutations. The metrics for each dataset are calculated and the values represent the average across all datasets.)

| Model | All mutations | | | | Single-point mutations | | | |
|---|---|---|---|---|---|---|---|---|
| | Spearman | NDCG | Acc | F1 | Spearman | NDCG | Acc | F1 |
| CARP(640M) | 0.167 | 0.850 | 0.578 | 0.523 | 0.201 | 0.863 | 0.596 | 0.543 |
| ESM IF1 | **0.215** | 0.854 | 0.598 | 0.535 | 0.251 | 0.867 | 0.611 | 0.551 |
| ESM1b | 0.152 | 0.851 | 0.587 | 0.523 | 0.200 | 0.865 | 0.604 | 0.543 |
| ESM1v(ensemble) | 0.158 | 0.850 | 0.577 | 0.525 | 0.200 | 0.863 | 0.594 | 0.546 |
| ESM2(650M) | 0.167 | 0.851 | 0.583 | 0.528 | 0.207 | 0.864 | 0.598 | 0.545 |
| GEMME | 0.206 | 0.857 | 0.597 | 0.537 | 0.258 | **0.872** | 0.613 | 0.552 |
| MIF | **0.229** | **0.859** | **0.612** | **0.547** | **0.261** | **0.872** | **0.627** | **0.562** |
| MIFST | 0.213 | 0.858 | **0.607** | **0.539** | 0.240 | 0.870 | 0.622 | 0.556 |
| MSA Transformer(ensemble) | 0.175 | 0.856 | 0.588 | 0.528 | 0.219 | 0.869 | 0.604 | 0.545 |
| MULAN(small) | 0.110 | 0.840 | 0.569 | 0.526 | 0.137 | 0.851 | 0.577 | 0.537 |
| PoET | 0.202 | 0.858 | 0.594 | 0.538 | 0.244 | 0.870 | 0.611 | 0.555 |
| VenusREM | **0.23** | **0.863** | **0.605** | **0.54** | **0.285** | **0.878** | **0.623** | **0.558** |
| ProSSN(k=20,h=512) | 0.149 | 0.852 | 0.589 | 0.53 | 0.199 | 0.866 | 0.603 | 0.549 |
| ProSST(2048) | 0.208 | **0.859** | 0.603 | 0.537 | **0.259** | **0.874** | 0.620 | 0.556 |
| Progen2(large) | 0.137 | 0.843 | 0.581 | 0.524 | 0.168 | 0.855 | 0.599 | 0.543 |
| ProtGPT2 | 0.058 | 0.825 | 0.556 | 0.507 | 0.065 | 0.829 | 0.555 | 0.506 |
| ProteinMPNN | 0.133 | 0.838 | 0.573 | 0.528 | 0.157 | 0.848 | 0.577 | 0.535 |
| RITA(xlarge) | 0.117 | 0.842 | 0.572 | 0.515 | 0.154 | 0.855 | 0.588 | 0.534 |
| SaProt(650M-PDB) | 0.202 | 0.855 | 0.588 | 0.524 | 0.237 | 0.866 | 0.603 | 0.542 |
| Tranception L | 0.181 | 0.854 | 0.592 | 0.539 | 0.222 | 0.867 | 0.608 | **0.558** |
| Tranception L(no retrieval) | 0.114 | 0.840 | 0.57 | 0.519 | 0.143 | 0.851 | 0.584 | 0.535 |
| UniRep | -0.055 | 0.806 | 0.495 | 0.478 | -0.059 | 0.810 | 0.491 | 0.471 |
| VESPA | 0.163 | 0.853 | 0.584 | 0.517 | 0.216 | 0.867 | 0.600 | 0.531 |

**First**, **Second**, **Third** best performance for each metric.

## 3.2. Analysis by functional categories

### 3.2.1. Stability

Stability predictions, comprising the largest subset of our benchmark ($N$=498, 62.5% of total datasets), revealed clear patterns in model performance that align with our understanding of protein stability mechanisms. Given that protein stability is inherently linked to structural characteristics, it is notable that structure aware models, particularly inverse folding models, demonstrated superior predictive capabilities in this category. MIF emerged as the top performer across all evaluation metrics (Table 2), achieving a Spearman correlation of 0.369 , NDCG of 0.898 , accuracy of 0.620, and F1 score of 0.599. Other structure-aware models also showed strong performance, with ESM-IF1 and MIFST achieving Spearman correlations of 0.343 and 0.336 respectively. Furthermore, VenusREM's strong performance (Spearman correlation 0.355, NDCG 0.897, accuracy 0.613) demonstrates that integrating multiple information sources (structural, evolutionary, and sequence information) can provide complementary signals for stability assessment, achieving the second-best performance across most metrics.

Table 2: Model performance on stability-related mutations

| Model | Spearman | NDCG | Acc | F1 |
|---|---|---|---|---|
| CARP(640M) | 0.242 | 0.882 | 0.582 | 0.565 |
| ESM IF1 | **0.343** | 0.893 | 0.610 | **0.589** |
| ESM1b | 0.223 | 0.878 | 0.584 | 0.562 |
| ESM1v(ensemble) | 0.232 | 0.881 | 0.578 | 0.567 |
| ESM2(650M) | 0.240 | 0.88 | 0.583 | 0.559 |
| GEMME | 0.300 | 0.887 | 0.594 | 0.574 |
| MIF | **0.369** | **0.898** | **0.620** | **0.599** |
| MIFST | 0.336 | 0.893 | **0.612** | **0.588** |
| MSA Transformer(ensemble) | 0.255 | 0.885 | 0.588 | 0.570 |
| MULAN(small) | 0.179 | 0.873 | 0.566 | 0.558 |
| PoET | 0.271 | 0.887 | 0.597 | 0.575 |
| **VenusREM** | **0.355** | **0.897** | **0.613** | **0.584** |
| ProSSN(k=20,h=512) | 0.227 | 0.881 | 0.582 | 0.567 |
| ProSST(2048) | 0.332 | **0.894** | 0.607 | 0.582 |
| Progen2(large) | 0.196 | 0.870 | 0.578 | 0.559 |
| ProtGPT2 | 0.077 | 0.847 | 0.538 | 0.521 |
| ProteinMPNN | 0.236 | 0.879 | 0.583 | 0.572 |
| RITA(xlarge) | 0.178 | 0.870 | 0.565 | 0.549 |
| SaProt(650M-PDB) | 0.283 | 0.885 | 0.591 | 0.565 |
| Tranception L | 0.256 | 0.882 | 0.590 | 0.570 |
| Tranception L(no retrieval) | 0.175 | 0.866 | 0.560 | 0.542 |
| UniRep | -0.088 | 0.828 | 0.488 | 0.470 |
| VESPA | 0.212 | 0.876 | 0.572 | 0.539 |

**First**, **Second**, **Third** best performance for each metric.

### 3.2.2. Activity

In activity predictions ($N$=130, 16.3% of total datasets), we observed distinct performance patterns across different model types. VESPA, which combines conservation predictions (from ProtT5 embeddings), BLOSUM62 substitution scores, and context-aware substitution probabilities, achieved the highest performance in ranking metrics (Table 3), with a Spearman correlation 0.338 and NDCG of 0.833. Models directly utilizing MSA information, such as GEMME (Spearman correlation= 0.277, NDCG= 0.828), Tranception (Spearman correlation= 0.273), and VenusREM (Spearman correlation= 0.291, NDCG= 0.831), also demonstrated strong predictive power across different evaluation metrics. The competitive performance of these evolution-informed approaches, including PoET (Spearman correlation= 0.302) which utilizes homologous sequence information, suggests that evolutionary information, whether used directly or indirectly, is particularly valuable for predicting activity-altering mutations. Meanwhile, inverse folding models achieved the highest accuracy scores, with MIFST and MIF reaching accuracies of 0.721 and 0.717 respectively, while Tranception excelled in identifying beneficial mutations as indicated by its top F1 score of 0.559. This diverse performance pattern suggests that different modeling approaches may capture complementary aspects of protein activity mechanisms. The superior accuracy of inverse folding models indicates their strength in structure-aware binary classification, while evolution-informed models excel at capturing the nuanced effects of mutations through ranking metrics. This observation highlights the potential benefit of integrating both structural and evolutionary information for activity prediction, as demonstrated by the balanced performance of models like VenusREM across different metrics (Spearman correlation= 0.291, NDCG= 0.831).

Table 3: Model performance on activity-related mutations

| Model | Spearman | NDCG | Acc | F1 |
|---|---|---|---|---|
| CARP(640M) | 0.250 | 0.806 | 0.661 | 0.504 |
| ESM IF1 | 0.217 | 0.802 | 0.685 | 0.516 |
| ESM1b | 0.249 | 0.822 | 0.709 | 0.531 |
| ESM1v(ensemble) | **0.291** | 0.814 | 0.667 | 0.512 |
| ESM2(650M) | 0.275 | 0.815 | 0.679 | 0.513 |
| GEMME | 0.277 | **0.828** | **0.712** | **0.542** |
| MIF | 0.172 | 0.799 | **0.717** | 0.524 |
| MIFST | 0.151 | 0.802 | **0.721** | 0.531 |
| MSA Transformer(ensemble) | 0.266 | 0.821 | 0.693 | 0.517 |
| MULAN(small) | 0.150 | 0.770 | 0.642 | 0.498 |
| PoET | **0.302** | 0.820 | 0.690 | 0.526 |
| VenusREM | 0.291 | **0.831** | 0.708 | 0.532 |
| ProSSN(k=20,h=512) | 0.236 | 0.817 | 0.707 | **0.537** |
| ProSST(2048) | 0.249 | 0.823 | 0.711 | 0.531 |
| Progen2(large) | 0.228 | 0.805 | 0.682 | 0.521 |
| ProtGPT2 | 0.078 | 0.735 | 0.598 | 0.452 |
| ProteinMPNN | 0.026 | 0.73 | 0.577 | 0.444 |
| RITA(xlarge) | 0.22 | 0.802 | 0.679 | 0.515 |
| SaProt(650M-PDB) | 0.233 | 0.808 | 0.684 | 0.511 |
| Tranception L | 0.273 | 0.818 | 0.705 | **0.559** |
| Tranception L(no retrieval) | 0.180 | 0.793 | 0.673 | 0.525 |
| UniRep | -0.104 | 0.680 | 0.482 | 0.407 |
| VESPA | **0.338** | **0.833** | 0.708 | 0.531 |

**First**, **Second**, **Third** best performance for each metric.

### 3.2.3. Binding

We separated binding affinity predictions into protein–protein interactions (PPI, $N$=100, 12.6% of total datasets) and drug-target interactions (DTI, $N$=42, 5.3% of total datasets), considering their distinct molecular mechanisms and interaction interfaces (Table 4). This separation revealed interesting patterns in model performance across different binding types. In PPI predictions, most models showed relatively modest performance, suggesting the inherent complexity of protein–protein binding interfaces. Notably, the multichain versions of ESM-IF1 and ProteinMPNN significantly outperformed their single-chain counterparts: ESM-IF1 (multichain) achieved Spearman correlation of 0.193 and NDCG of 0.875 compared to its single-chain version (Spearman correlation=0.049, NDCG=0.856), while ProteinMPNN (multichain) showed similar improvements (Spearman correlation 0.130 vs 0.034), demonstrating the importance of incorporating partner protein structural information for accurate PPI predictions. VESPA also achieved competitive performance across all metrics (Spearman correlation=0.148, NDCG=0.881, accuracy=0.572, F1 score=0.551), indicating that the combination of conservation patterns and substitution probabilities can effectively capture protein–protein binding effects even without explicit partner information.

For DTI predictions, models generally demonstrated better performance, with less pronounced differences between different modeling approaches. The top performers included structure-aware SaProt (Spearman correlation 0.271), MSA-based GEMME (Spearman correlation 0.240), and PoET (NDCG 0.887), which utilizes homologous sequences. All these models achieved competitive results across different metrics, with PoET notably reaching an accuracy of 0.756 and F1 score of 0.623. This balanced performance suggests that drug-protein binding might be more effectively captured by current computational approaches, possibly due to the more localized nature of protein-small molecule interactions.

Table 4: Model performance on binding-related mutations

| | PPI Binding | | | | DTI Binding | | | |
|---|---|---|---|---|---|---|---|---|
| **Model** | **Spearman** | **NDCG** | **Acc** | **F1** | **Spearman** | **NDCG** | **Acc** | **F1** |
| CARP(640M) | 0.006 | 0.850 | 0.542 | 0.513 | 0.178 | 0.878 | 0.742 | 0.577 |
| ESM IF1 | 0.049 | 0.856 | 0.548 | 0.503 | -0.07 | 0.817 | 0.622 | 0.467 |
| ESM IF1(multichain) | **0.193** | **0.875** | **0.573** | **0.537** | - | - | - | - |
| ESM1b | 0.072 | 0.869 | 0.544 | 0.524 | 0.217 | 0.884 | 0.748 | 0.588 |
| ESM1v(ensemble) | 0.003 | 0.85 | 0.537 | 0.519 | 0.177 | 0.882 | 0.751 | 0.591 |
| ESM2(650M) | 0.019 | 0.858 | 0.538 | 0.536 | 0.204 | **0.886** | 0.730 | 0.608 |
| GEMME | 0.102 | 0.870 | 0.558 | 0.531 | **0.24** | 0.885 | 0.752 | 0.592 |
| MIF | 0.013 | 0.857 | 0.539 | 0.501 | 0.077 | 0.875 | 0.745 | 0.578 |
| MIFST | 0.029 | 0.861 | 0.532 | 0.499 | 0.131 | 0.874 | 0.752 | 0.585 |
| MSA | 0.077 | 0.867 | 0.544 | 0.514 | 0.145 | 0.88 | 0.755 | 0.592 |
| Transformer(ensemble) | | | | | | | | |
| MULAN(small) | -0.016 | 0.85 | 0.517 | 0.512 | 0.164 | 0.872 | 0.730 | 0.559 |
| PoET | 0.118 | 0.868 | 0.552 | 0.525 | 0.200 | **0.887** | **0.756** | **0.623** |
| VenusREM | 0.069 | 0.864 | 0.553 | 0.526 | 0.164 | 0.879 | 0.752 | 0.593 |
| ProSSN(k=20,h=512) | 0.065 | 0.866 | 0.553 | **0.542** | 0.166 | 0.882 | 0.748 | 0.585 |
| ProSST(2048) | 0.063 | 0.860 | 0.552 | 0.52 | 0.104 | 0.868 | 0.747 | 0.579 |
| Progen2(large) | 0.049 | 0.856 | 0.559 | 0.523 | 0.073 | 0.867 | 0.748 | 0.59 |
| ProtGPT2 | 0.071 | 0.869 | **0.561** | 0.531 | -0.144 | 0.817 | 0.607 | 0.447 |
| ProteinMPNN | 0.034 | 0.857 | 0.535 | 0.495 | 0.038 | 0.839 | 0.621 | 0.498 |
| ProteinMPNN(multichain) | **0.13** | **0.874** | 0.547 | 0.495 | - | - | - | - |
| RITA(xlarge) | 0.059 | 0.859 | 0.544 | 0.507 | 0.048 | 0.870 | 0.749 | 0.589 |
| SaProt(650M-PDB) | 0.049 | 0.852 | 0.531 | 0.521 | **0.271** | **0.886** | 0.742 | 0.561 |
| Tranception L | 0.057 | 0.865 | 0.541 | 0.521 | 0.176 | 0.881 | **0.763** | **0.622** |
| Tranception L(no retrieval) | 0.000 | 0.85 | 0.538 | 0.506 | 0.091 | 0.873 | **0.758** | **0.617** |
| UniRep | 0.099 | 0.867 | 0.536 | 0.524 | -0.052 | 0.837 | 0.406 | 0.473 |
| VESPA | **0.148** | **0.881** | **0.572** | **0.551** | **0.221** | 0.883 | 0.752 | 0.592 |

**First**, **Second**, **Third** best performance for each metric.

### 3.2.4. Selectivity

Selectivity predictions ($N$=26, 3.3% of total datasets), consisting of enantioselectivity (ee) and diastereoselectivity (de) measurements, presented the most challenging category in our evaluation. The performance across all models was notably poor, with most models achieving near-zero or even negative Spearman correlations (Table 5). Even the best performing models showed only marginally better results than random predictions: SaProt achieved the highest Spearman correlation of 0.099, while ProtGPT2 and ProSSN followed with correlations of 0.087 and 0.078 respectively. The classification metrics were similarly disappointing, with the highest accuracy of 0.570 (ProteinMPNN) and F1 score of 0.579 (UniRep) being substantially lower than those observed in other prediction tasks.

This universally poor performance can be attributed to several fundamental limitations of current prediction paradigms. First, selectivity measurements inherently involve the relative preference between different substrates (such as different enantiomers or diastereomers), while current models only consider protein information without explicitly modeling substrate-specific interactions. The ee and de values are derived from the relative concentrations of different products, making it particularly challenging for models to predict such differential binding or catalytic preferences without access to small molecule information. Second, the stereochemical nature of selectivity requires understanding of precise three-dimensional arrangements and chemical interactions, which current sequence-based and even structure-based models may not capture adequately.

To address these limitations, future improvements in selectivity prediction may require several key developments: (1) integration of protein-ligand docking simulations to explicitly model substrate binding modes, (2) incorporation of stereochemistry-aware features that can capture the spatial relationships determining enantioselectivity and diastereoselectivity, and (3) development of specialized architectures that can process both protein and substrate information in a manner that accounts for their relative spatial orientations. While the small dataset size (*N*=26) poses an additional challenge, the fundamental issue appears to be the mismatch between current modeling approaches and the inherent nature of selectivity prediction tasks.

Table 5: Model performance on selectivity-related mutations

| Model | Spearman | NDCG | Acc | F1 |
| --- | --- | --- | --- | --- |
| CARP(640M) | -0.019 | 0.81 | 0.506 | 0.367 |
| ESM IF1 | 0.041 | **0.837** | 0.508 | 0.337 |
| ESM1b | 0.016 | 0.812 | 0.476 | 0.242 |
| ESM1v(ensemble) | -0.029 | 0.809 | 0.484 | 0.339 |
| ESM2(650M) | -0.015 | 0.809 | 0.507 | 0.374 |
| GEMME | 0.038 | 0.809 | 0.469 | 0.228 |
| MIF | -0.041 | 0.815 | 0.486 | 0.253 |
| MIFST | -0.100 | 0.801 | 0.471 | 0.238 |
| MSA Transformer(ensemble) | 0.010 | 0.812 | 0.487 | 0.275 |
| MULAN(small) | -0.122 | 0.799 | 0.451 | 0.382 |
| PoET | 0.041 | 0.813 | 0.482 | 0.342 |
| VenusREM | -0.025 | 0.81 | 0.479 | 0.286 |
| ProSSN(k=20,h=512) | **0.078** | 0.822 | 0.483 | 0.25 |
| ProSST(2048) | -0.052 | 0.81 | 0.483 | 0.294 |
| Progen2(large) | -0.026 | 0.811 | 0.494 | 0.34 |
| ProtGPT2 | **0.087** | **0.834** | **0.555** | **0.476** |
| ProteinMPNN | 0.005 | 0.82 | **0.57** | **0.478** |
| RITA(xlarge) | -0.029 | 0.805 | 0.493 | 0.352 |
| SaProt(650M-PDB) | **0.099** | 0.826 | 0.498 | 0.32 |
| Tranception L | 0.067 | 0.827 | 0.502 | 0.389 |
| Tranception L(no retrieval) | 0.006 | 0.823 | 0.513 | 0.421 |
| UniRep | 0.055 | **0.828** | **0.538** | **0.579** |
| VESPA | -0.01 | 0.812 | 0.477 | 0.225 |

**First**, **Second**, **Third** best performance for each metric.

### *3.3. Impact of dataset size on model performance*

To investigate how dataset size affects model performance in protein mutation effect prediction, we analyzed 23 representative models using 905 mutation datasets. To ensure balanced sample distribution, we divided the datasets into four quartile-based groups based on mutation counts: 5–8 mutations (234 datasets), 8–13 mutations (236 datasets), 13–28 mutations (215 datasets), and >28 mutations (220 datasets).

As shown in Fig. 3, we categorized models into three groups based on their underlying principles: structure-aware models, sequence-only models, and evolution-informed models. Structure-aware models generally demonstrated strong performance scaling, with VenusREM and MIF achieving the highest correlations ($\rho \approx 0.36$ and $\rho \approx 0.35$) on large datasets (>28 mutations). Notably, ESM-IF1 showed comparable performance ($\rho \approx 0.34$), suggesting the effectiveness of structure-based approaches. In the sequence-only category, ESM2-650M emerged as the top performer ($\rho \approx 0.29$), followed by ESM1v and ESM1b, though some models like UniRep showed limited effectiveness ($\rho < 0$). Evolution-informed models displayed remarkable consistency, with VenusREM achieving the highest correlation ($\rho \approx 0.36$) and other models (GEMME, MSA Transformer, PoET) reaching $\rho \approx 0.30$ on large datasets.

These findings suggest that model selection should consider both the available experimental data size and the desired prediction approach. Structure-aware and evolution-informed models generally provide more reliable predictions across different dataset sizes, while sequence-only models show greater performance variation and may require larger datasets ($>13$ mutations) for optimal performance. The most substantial improvements typically occur between the 5–8 and 8–13 mutation groups, indicating a minimum threshold for reliable predictions.

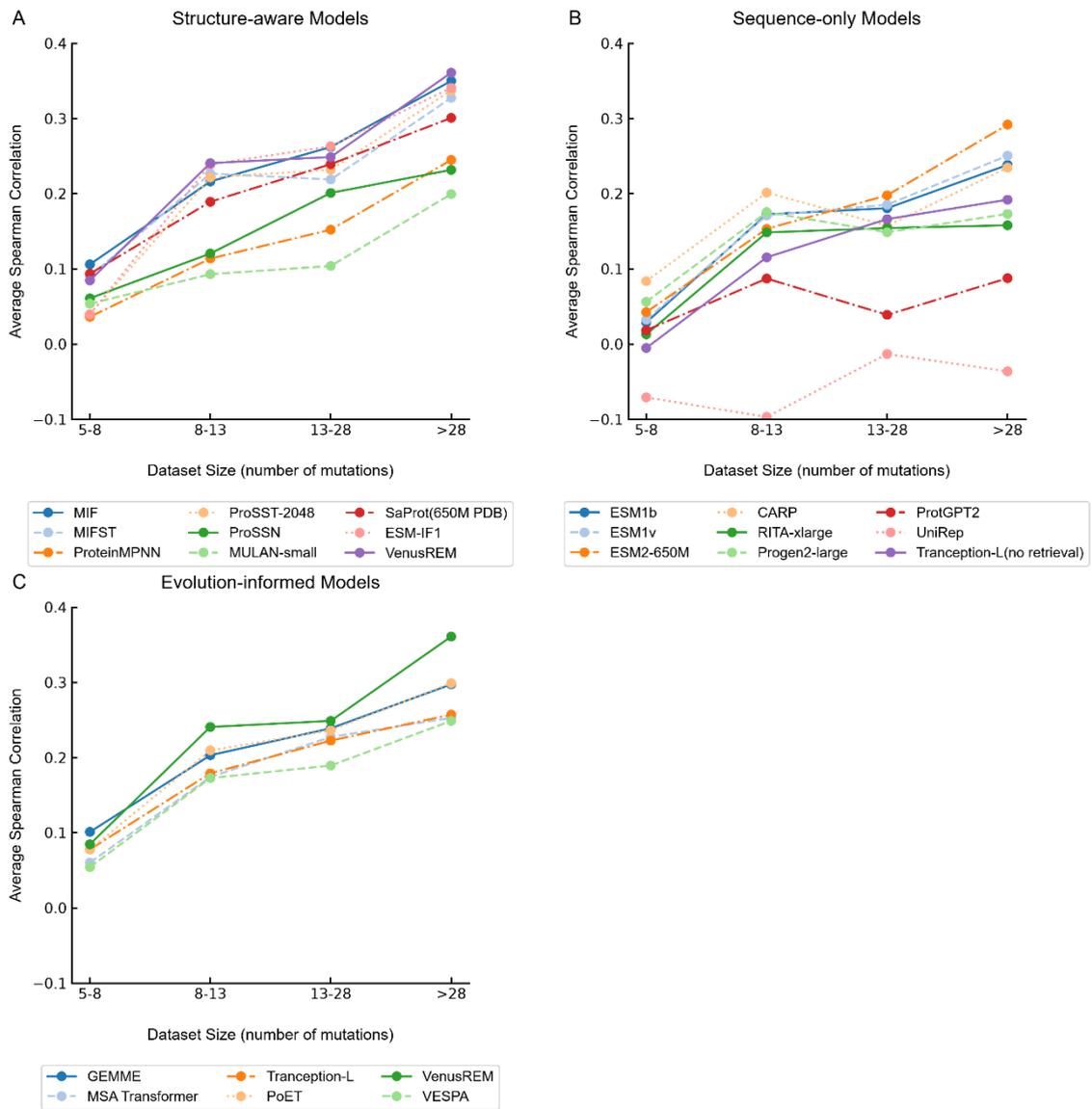

**Figure 3** Impact of dataset size on model performance. Models are categorized into structure-aware, sequence-only, and evolution-informed approaches. The average Spearman correlation coefficient ($\rho$) is calculated across all mutation datasets within each size group.

### 3.4. Model performance variance analysis

In addition to average performance metrics, the consistency of model predictions across different datasets is crucial for practical applications. Fig. 4 shows the average standard deviation of Spearman correlation coefficients for different model types across functional categories, providing insights into prediction stability.

Our analysis reveals significant variations in prediction stability across different model types and functional categories. For stability predictions, structure-aware models demonstrate lower variance ($\sigma \approx 0.35$) compared to sequence-only models ($\sigma \approx 0.39$), suggesting that incorporating structural information leads to more reliable stability predictions across diverse proteins. In contrast, for activity predictions, evolution-informed models show more consistent performance ($\sigma \approx 0.29$) compared to both structure-aware ($\sigma \approx 0.32$) and sequence-only models ($\sigma \approx 0.30$). This pattern indicates that evolutionary information provides more stable signals for activity-related mutations across different protein families. The highest variance was observed in selectivity predictions for both evolution-informed and structure-aware models ($\sigma \approx 0.51$), highlighting the particularly challenging nature of this prediction task. Notably, sequence-only models showed surprisingly lower variance in selectivity predictions ($\sigma \approx 0.25$), though this may be influenced by their generally lower performance overall. For binding affinity predictions, structure-aware models exhibited lower variance in protein–protein interactions ($\sigma \approx 0.30$) compared to evolution-informed models ($\sigma \approx 0.36$), indicating that structural information provides more consistent signals for PPI predictions. However, for drug-target interactions, sequence-only models showed competitive consistency ($\sigma \approx 0.32$) compared to structure-aware models ($\sigma \approx 0.39$).

These findings emphasize that model selection should consider not only average performance but also prediction consistency. In practical applications where reliability is paramount, models with lower performance variance may be preferable even if they do not achieve the highest average scores. Furthermore, this analysis highlights the need for uncertainty quantification methods in mutation effect prediction, which would allow practitioners to make more informed decisions by considering both the predicted effect and the confidence level of that prediction.

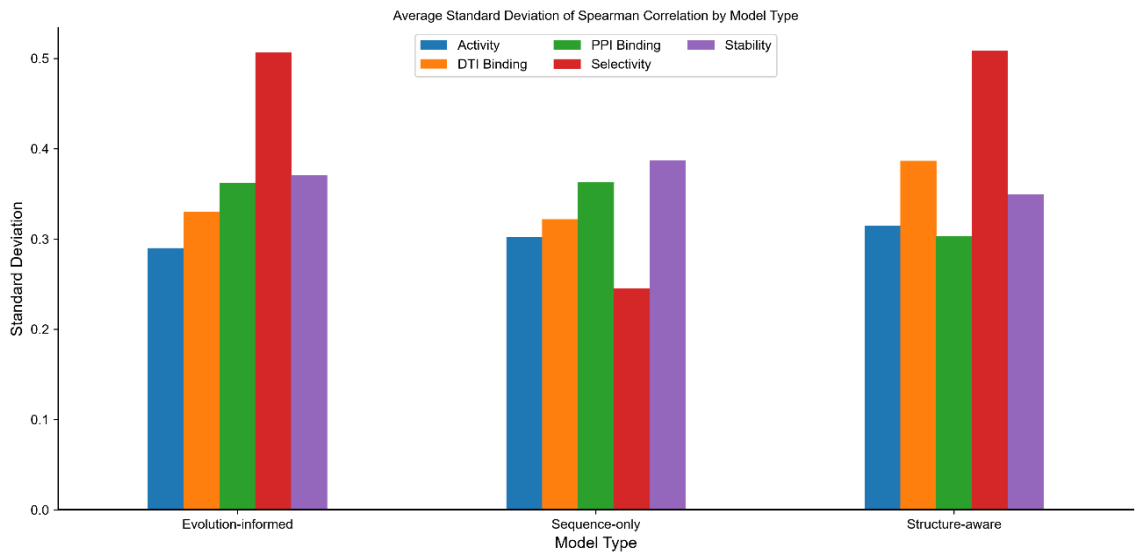

**Figure 4** Average standard deviation of Spearman correlation by model type across different functional categories. Lower standard deviation indicates more consistent predictions across different datasets within each category.

## 4. Conclusions

In this study, we constructed a comprehensive mutation benchmark dataset encompassing 905 distinct datasets across stability, activity, binding, and selectivity properties, and systematically evaluated 23 protein mutation effect prediction models. Our analysis revealed distinct performance patterns: structure-aware models, particularly inverse folding models like MIF and MIFST, demonstrated superior performance in stability predictions, while evolution-informed models like VESPA and GEMME excelled in activity predictions. VenusREM, which integrates sequence, structure, and evolutionary information, showed strong overall performance across multiple metrics. For binding affinity predictions, we observed different patterns between protein–protein interactions and drug-target interactions, with models incorporating partner protein structural information showing advantages in PPI predictions. Notably, all models showed substantially better performance on single-point mutations compared to multiple mutations, and selectivity predictions emerged as the most challenging category.

Our analysis of dataset size impact further revealed that model performance generally improves with larger datasets, with structure-aware and evolution-informed models providing more reliable predictions across different dataset sizes. We identified a critical threshold around 8–13 mutations, where most models show substantial performance improvements, suggesting a minimum data requirement for reliable predictions. Additionally, our performance variance analysis demonstrated that different model types exhibit varying levels of prediction consistency across functional categories, with structure-aware models showing more reliable stability predictions, while evolution-informed models provide more consistent activity predictions. These findings highlight that model selection should consider not only average performance but also prediction consistency and the available experimental data size.

These findings have significant implications for both protein engineering and pharmaceutical science. In biopharmaceutical development, the superior performance of structure-aware models in stability prediction and the distinct patterns in drug-target interaction predictions can guide the selection of appropriate computational tools in different stages of drug discovery pipelines. The identified challenges in predicting protein–protein interactions and the advantages of multichain models highlight important considerations for developing therapeutic antibodies and other protein-based drugs. However, several limitations should be considered: the uneven distribution across different protein properties (stability-related mutations comprising 62.5% while selectivity-related only 3.3%), potential sampling bias in protein families, and the dataset's composition might not fully reflect the diverse challenges in practical applications.

Looking forward, advancing mutation effect prediction will require both architectural innovations and better integration of different information sources. The complementary strengths of different modeling approaches suggest that hybrid or ensemble methods might offer promising directions for future development, particularly for challenging properties like multiple mutations, selectivity, and protein–protein interactions. Furthermore, incorporating uncertainty quantification methods would enhance the practical utility of these models by providing reliability measures alongside predictions. These insights can help both protein engineers and pharmaceutical scientists make more informed decisions in their computational workflows, especially in the early stages of development where experimental validation is costly and time-consuming.

**Data availability**

The protein mutation dataset used in this study is publicly available at the Hugging Face repository (https://huggingface.co/datasets/AI4Protein/VenusMutHub).

**Author Contributions**

Liang Zhang: Writing - original draft, Software, Visualization, Methodology, Data curation. Hua Pang: Data curation. Chenghao Zhang: Data curation, Visualization. Song Li: Resources. Yang Tan: Resources. Fan Jiang: Resources. Yuanxi Yu: Resources. Mingchen Li: Methodology. Ziyi Zhou: Methodology, Review. Banghao Wu: Methodology. Bingxin Zhou: Methodology. Pan Tan: Writing - review & editing, Project administration, Funding acquisition, Supervision. Liang Hong:

Writing -review & editing, Project administration, Funding acquisition, Supervision.

**Acknowledgments**

This work was supported by Science and Technology Innovation Key R&D Program of Chongqing(CSTB2024TIAD-STX0032), the Computational Biology Key Program of Shanghai Science and Technology Commission (23JS1400600), Shanghai Jiao Tong University Scientific and Technological Innovation Funds (21X010200843), and Science and Technology Innovation Key R&D Program of Chongqing (CSTB2022TIAD-STX0017), the Postdoctoral Fellowship Program of CPSF under Grant Number GZC20241010, the Student Innovation Center at Shanghai Jiao Tong University, and Shanghai Artificial Intelligence Laboratory

**Conflicts of interest**

The authors declare no potential conflicts of interest.